\documentclass[pra,aps,twocolumn,showpacs]{revtex4}

\usepackage{bm}
\usepackage{epsfig}

\begin{document}


\title{Optimal partial estimation of quantum states from several copies}
\author{Ladislav Mi\v{s}ta Jr.}  
\affiliation{Department of Optics, Palack\' y University, 17.
listopadu 50,  772~07 Olomouc, Czech Republic}
\author{Jarom\'{\i}r Fiur\'{a}\v{s}ek}  
\affiliation{Department of Optics, Palack\' y University, 17.
listopadu 50,  772~07 Olomouc, Czech Republic}
\date{\today}

\begin{abstract}
We derive analytical formula for the optimal trade-off between the mean estimation 
fidelity and the mean fidelity of the qubit state after a partial measurement on 
$N$ identically prepared qubits. We also conjecture 
analytical expression for  the optimal fidelity trade-off in case of
a partial measurement on $N$ identical copies of a $d$-level system. 
\end{abstract}

\maketitle

\section{Introduction}

Certain operations permitted in classical physics cannot be 
done perfectly in quantum mechanics. This is best exemplified by the 
celebrated no-cloning theorem \cite{Wootters_82} which forbids to create 
two perfect copies of an unknown quantum state. The no-cloning theorem is 
closely related to 
another no-go theorem stating that one cannot gain some information on 
an unknown  quantum state without disturbing it. Namely, if this would 
be possible one would be able to prepare two approximate replicas of this state 
which would be better than the best ones allowed by quantum mechanics 
\cite{Niu_98,Cerf98,Buzek98,Cerf00}. Therefore, in quantum mechanics any operation on an 
unknown quantum state giving some information on the state inevitably 
disturbs the state and in addition, the more information it extracts the 
larger is the disturbance. This fundamental property of quantum operations is 
reflected in the plane of values of quantities quantifying the information 
gain and the state disturbance by a certain optimal trade-off curve that 
cannot be overcome by any quantum operation. Among all quantum operations 
particularly interesting are those which lie on this curve since they to 
the best possible extent as quantum mechanics allows approximate 
ideal disturbance-free measurement device. These operations, conventionally 
denoted as minimal disturbance measurements (MDMs), in general depend on 
the set of input states, their a priori distribution and also on the 
quantities quantifying the information gain and the state disturbance 
\cite{Banaszek_05}. The most successful approach to the finding of the optimal 
trade-offs and corresponding MDMs proved to be that based on the quantification 
of the information gain by the mean estimation fidelity $G$ and the state disturbance 
by the mean output fidelity $F$ \cite{Banaszek_01a}. Using this approach 
it was possible to derive analytically the optimal trade-offs between $G$ and $F$ 
and  to find the MDMs for a single copy of a completely unknown pure state 
of a $d$-level system \cite{Banaszek_01a}, a completely unknown pure state 
of a $d$-level system produced by $d$ independent phase shifts of some 
reference state  \cite{Mista_05} and a completely unknown maximally 
entangled state of two $d$-level systems \cite{Sacchi_06}.  
Besides, the first two above mentioned MDMs were also demonstrated experimentally for 
$d=2$ (qubit) \cite{Sciarrino_06}. The studies on MDM were not restricted to finite-dimensional 
quantum systems and also MDM for a completely unknown coherent state was found 
and realized experimentally in \cite{Andersen_06}. Multi-copy MDMs were introduced by Banaszek and Devetak  who 
considered  partial measurement on $N$ identical copies of a pure qubit state. 
They assumed MDMs which output $N$ disturbed quantum copies of the state 
and a classical estimate and they {\it numerically}
found the optimal fidelity trade-off 
for this scenario \cite{Banaszek_01b}. 
The MDMs are not only of fundamental importance but they can be also applied to increase 
transmission fidelity of certain lossy and noisy channels \cite{Ricci_05,Andersen_06}.

In this paper we further investigate the minimum disturbance measurement on
several copies of the state. In contrast to Ref. \cite{Banaszek_01b} we assume
operations which output only a \emph{single} quantum copy of the input state.
We derive \emph{analytically} the optimal trade-off between the mean 
estimation fidelity $G$ and the mean output fidelity $F$ for an ensemble of 
$N$ identical pure  qubits which is given by the formula 
\begin{equation}\label{Nqubit}
\sqrt{F-\frac{1}{N+2}}= \sqrt{\frac{N+1}{N+2}-G}+\sqrt{N\left(G-\frac{N}{N+2}\right)}.
\end{equation}
Moreover, we also conjecture that the optimal fidelity trade-off for an ensemble of 
$N$ identical pure states of a $d$-level system has the form
\begin{eqnarray}\label{Nqudit}
\sqrt{F-\frac{1}{N+d}}&=&\sqrt{(d-1)\left(\frac{N+1}{N+d}-G\right)}\nonumber\\
&&+\sqrt{N\left(G-\frac{N}{N+d}\right)}.
\end{eqnarray}

The paper is organized as follows. The general formalism allowing to determine 
the MDM is presented in Sec.~\ref{sec_1}. 
In Sec.~\ref{sec_qubits} we find the optimal fidelity trade-off and 
the corresponding MDM for $N$ identical qubits. In Sec.~\ref{sec_2} we present a conjecture of 
the optimal fidelity trade-off for $N$ identical $d$-level systems. Finally, Sec.~\ref{sec_3} 
contains conclusions.

\section{Minimal disturbance measurement}\label{sec_1}

Let us investigate  a  general MDM for $N$ identical pure states of 
a $d$-level system (qudit). Such states are represented by vectors 
in a $d$-dimensional Hilbert space $\mathcal{H}^{(d)}$ with an orthonormal 
basis $\{|0\rangle,|1\rangle,\ldots,|d-1\rangle\}$. The qudits form 
an orbit of the group  $\mathrm{SU}(d)$ of $d\times d$ unitary matrices 
with determinant $+1$, $|\psi(g)\rangle=U_{d}(g)|0\rangle$, 
where $U_{d}(g)$, $g\in \mathrm{SU}(d)$, is a unitary representation of $\mathrm{SU}(d)$ 
on $\mathcal{H}^{(d)}$. 
We consider here quantum operations on $N$ identical qudits
\begin{equation}\label{Nquditstate}
|\psi(g)\rangle^{\otimes N}=[U_{d}(g)]^{\otimes N}|0\rangle^{\otimes N}.
\end{equation}
The operation   outputs a {\it single} qudit - an approximate replica of
$|\psi(g)\rangle$ - and also yields a classical estimate of 
$|\psi(g)\rangle$. Without loss of generality, these  estimates 
can be labeled by the elements of the group $\mathrm{SU}(d)$.
Note that the input Hilbert space of the operation is the symmetric subspace 
of the Hilbert space of $N$ qudits, $\mathcal{H}_{\mathrm{in}}=\mathcal{H}_{+,N}^{(d)}$
and the output Hilbert space is the space of a single qudit,
$\mathcal{H}_{\mathrm{out}}=\mathcal{H}^{(d)}$.

Our task is to find an operation which exhibits the best possible 
performance in the following protocol \cite{Banaszek_01a}. In each run of 
the protocol,  the operation is applied on the quantum state (\ref{Nquditstate}).
We assume that $|\psi(g)\rangle^{\otimes N}$ is 
chosen randomly with uniform a priori distribution from the set of states 
$\{|\psi(g)\rangle^{\otimes N}\}_{g\in \mathrm{SU}(d)}$. If the outcome 
$h\in \mathrm{SU}(d)$ 
is detected the operation produces a single qudit output state $\rho(h|g)$. 
This state is not normalized and its trace $P(h|g)\equiv\mathrm{Tr}[\rho(h|g)]$ is 
the probability density of obtaining the outcome $h$ on the state (\ref{Nquditstate}). 
The information on the state $|\psi(g)\rangle$ contained in the
measurement outcome $h$ is converted 
into a guess of the state which is in our case a single qudit state $|\psi(h)\rangle$. 
The performance of this procedure can be quantified by two mean fidelities: 
mean output fidelity $F$ defined as 
\begin{equation}\label{F}
F=\int_{\mathrm{SU}(d)}\int_{\mathrm{SU}(d)}\langle\psi(g)|\rho(h|g)|\psi(g)\rangle d h d g,
\end{equation}
which quatifies the average state disturbance and the mean estimation fidelity $G$ 
defined by the formula
\begin{equation}\label{G} 
G=\int_{\mathrm{SU}(d)}\int_{\mathrm{SU}(d)}P(h|g)|\langle\psi(g)|\psi(h)\rangle|^{2} d h d g,
\end{equation}
which quantifies the average information gain. Here, the integrals are taken over the 
whole group $\mathrm{SU}(d)$ and $dg$ is the normalized invariant Haar measure on the group.
Quantum mechanics sets a fundamental bound on the maximum value of the fidelity $F$ 
that can be attained for a given value of the fidelity $G$ for any considered quantum 
operation. The bound can be expressed in the form of a nontrivial optimal trade-off 
relation between $F$ and $G$ and the MDM is defined as a quantum operation for 
which the fidelities $G$ and $F$ satisfy the trade-off.  

Two extreme cases of the trade-off are well-known. First, if $G$ is the optimal 
estimation fidelity of the qudit state from $N$ identical copies, i.e. 
$G=(N+1)/(N+d)$ \cite{Bruss_99}, then $F$ can be at most equal to $F=(N+1)/(N+d)$. 
Second, if $F=1$ then $G$ cannot be larger than the optimal estimation fidelity 
of a qudit from $N-1$ identical copies, i.e. $G=N/(N-1+d)$. To find the whole 
optimal trade-off we can use the method developed in \cite{Mista_05}. With the help of
Jamiolkowski-Choi isomorphism \cite{Jamiolkowski72,Choi75} we can represent the completely positive map corresponding to
each particular outcome $h$ by a  positive-semidefinite operator $\chi_N^{(d)}(h)$
acting on the tensor product of the input and output Hilbert spaces
$\mathcal{H}_{+,N}^{(d)}\otimes \mathcal{H}^{(d)}$. It holds that 
$\rho(h|g)=\mathrm{Tr}_{\rm in}[\chi_{N}^{(d)}(h) (\psi^{\mathrm{T}}(g))^{\otimes N} \otimes \openone_{\mathrm{out}}]$
where $\psi(g)\equiv|\psi(g)\rangle\langle\psi(g)|$.  
As shown in \cite{Mista_05} the optimal partial measurement can  be assumed to be covariant 
which means that $\chi_N^{(d)}(h)$ are generated from a single properly 
normalized operator $\chi^{(d)}_{N}$,
\begin{equation}
\chi_{N}^{(d)}(h)= [U_d^{\ast \otimes N}(h) \otimes U_d(h)] \,\chi_N^{(d)} 
\,[U_d^{T \otimes N}(h) \otimes U_d^\dagger(h)].
\label{chicovariant}
\end{equation}
The overall operation must be trace-preserving which imposes the constraint,
\begin{equation}
\int_{\mathrm{SU}(d)} \mathrm{Tr}_{\mathrm{out}}[\chi_N^{(d)}(h)] d h = \openone_{\mathrm{in}},
\label{chinorm}
\end{equation}
where $\mathrm{Tr}_{\mathrm{out}}$ stands for the partial trace over the output single-qudit
Hilbert space and $\openone_{\mathrm{in}}$ denotes the identity operator on the input space
$\mathcal{H}_{+,N}^{(d)}$. 
The formula (\ref{chinorm}) expresses the completeness of the measurement carried on 
the input state. The unitary representation $U_d^{\otimes N}$ of $\mathrm{SU}(d)$ acts 
irreducibly on $\mathcal{H}_{+,N}^{(d)}$.  For the covariant map (\ref{chicovariant}) 
the integral  in Eq. (\ref{chinorm}) can  thus be evaluated with the help of 
Schur's lemma and we 
get $D(N,d)^{-1}\mathrm{Tr}[\chi_N^{(d)}]\openone_{\mathrm{in}}$ where 
$D(N,d)={N-1+d\choose d-1}$ is the dimension of the symmetric Hilbert space
$\mathcal{H}_{+,N}^{(d)}$. The trace-preservation condition (\ref{chinorm}) 
thus boils down to  the proper normalization of the map that should read 
$\mathrm{Tr}[\chi^{(d)}_{N}]=D(N,d)$.

The operator  $\chi^{(d)}_{N}$ generating the optimal partial measurement 
is proportional to a rank-one projector and can be written as 
$|\chi^{(d)}_{N}\rangle\langle\chi^{(d)}_{N}|$ where 
$|\chi^{(d)}_{N}\rangle$ is the eigenvector of a positive-semidefinite operator
\begin{equation}\label{Rp}
R^{(d)}_{p}=p R_{F}^{(d)}+(1-p)R_{G}^{(d)},\quad p\in[0,1]
\end{equation}
corresponding to its maximum eigenvalue \cite{Mista_05}. Here 
\begin{eqnarray}
R^{(d)}_{F}&=&\int_{\mathrm{SU}(d)}[\psi(g)^{\otimes N}]^{\rm T}\otimes \psi(g) dg,\label{RF}\\
R^{(d)}_{G}&=&\mathrm{Tr}_{\rm out}[R^{(d)}_F \openone_{\rm in} \otimes \psi(0)]\otimes\openone_{\rm out}\label{RG}.
\end{eqnarray}
Using the map $\chi^{(d)}_{N}$ the fidelities $F$ and $G$  can be expressed as
\begin{equation}\label{FG}
F=\mathrm{Tr}[\chi^{(d)}_{N}R^{(d)}_{F}], 
\qquad G=\mathrm{Tr}[\chi^{(d)}_{N}R^{(d)}_{G}].
\end{equation}
The operator $R^{(d)}_{F}$ can be easily evaluated using Schur's lemma and after some algebra 
we arrive at \cite{Fiurasek_04}
\begin{eqnarray}\label{RFd}
R^{(d)}_{F}&=&\frac{1}{D(N+1,d)}\left(\Pi^{(d)}_{+,N+1}\right)^{\rm T_{N}},
\end{eqnarray}
where $(\,\,\,)^{T_N}$ stands for the partial transposition with respect to the first $N$ 
qudits and $\Pi^{(d)}_{+,N+1}$ is the projector onto the subspace $\mathcal{H}_{+,N+1}^{(d)}$ . 
In what follows it is convenient to work with the occupation number basis 
\begin{eqnarray}\label{occupation}
|\{N_{i}\};N\rangle&=&|N_{0},N_{1},\ldots,N_{d-1};N\rangle=
\sqrt{\frac{N!}{\prod_{i=0}^{d-1}N_{i}!}}\nonumber\\
&\times&S_{N}|\underbrace{00\ldots0}_{N_{0}}\underbrace{11\ldots1}_{N_{1}}
\cdots\underbrace{d-1\ldots d-1}_{N_{d-1}}\rangle,\nonumber\\
\end{eqnarray}
that forms an orthonormal basis in the subspace $\mathcal{H}^{(d)}_{+,N}$. 
Here $S_{N}=(1/N!)\sum_{\{\pi\}}P_{\pi}^{(N)}$ is the symmetrization operator for $N$ 
qudits, the symbol $\{\pi\}$ stands for summation over all $N!$ permutations 
of $N$ qudits and $P_{\pi}^{(N)}$ denotes the permutation operator of $N$ qudits; 
the integers $N_{i}$, $N\geq N_{i}\geq 0$, $i=0,\ldots, d-1$ are the numbers of qudits 
in the states $|i\rangle$, $i=0,\ldots, d-1$ that satisfy the constraint 
$\sum_{i=0}^{d-1}N_{i}=N$. Making use of the occupation number basis the operator 
$\Pi^{(d)}_{+,N+1}$ can be expressed as 
\begin{eqnarray}\label{Pi}
\Pi^{(d)}_{+,N+1}=\!\!\!\sum_{\sum_{i=0}^{d-1}N_{i}=N+1}\!\!\!\!\!\!\!\!|\{N_{i}\};N+1\rangle\langle\{N_{i}\};N+1|. 
\end{eqnarray}    
To find the desired MDM for $N$ identical qudits we have to diagonalize a large 
matrix $R^{(d)}_{p}$. For a general $d$ this is a complex task which 
can be solved numerically. However, if we resort to the qubit case ($d=2$) we can find 
the optimal fidelity trade-off and the MDM analytically. The obtained result 
then can be used to make at least a conjecture about the optimal fidelity trade-off 
for $N$ qudits. 

\section{$N$ identical qubits}\label{sec_qubits}

For qubits the operator (\ref{Pi}) reads as
\begin{eqnarray}
\Pi^{(2)}_{+,N+1}=\sum_{k=0}^{N+1}|N+1,k\rangle\langle N+1,k|,
\end{eqnarray}
where $|N,k\rangle\equiv|N_{0}=N-k,N_{1}=k;N\rangle$ is a completely symmetric state of 
$N$ qubits in which $k$ qubits are in the basis state $|1\rangle$ and the remaining 
$N-k$ qubits are in the basis state $|0\rangle$. Hence, making use of the formula
\begin{eqnarray}
|N+1,k\rangle&=&\sqrt{\frac{N-k+1}{N+1}}|N,k\rangle|0\rangle\nonumber\\
&&+\sqrt{\frac{k}{N+1}}|N,k-1\rangle|1\rangle
\end{eqnarray}
and Eq.~(\ref{RFd}) one finds that
\begin{eqnarray}
R^{(2)}_{F}&=&\frac{1}{(N+1)(N+2)}\nonumber\\
&&\times\sum_{k=0}^{N+1}\left[(N-k+1)|N,k\rangle|0\rangle
\langle N,k|\langle0|\right.\nonumber\\
&&\left.+k|N,k-1\rangle|1\rangle\langle N,k-1|\langle 1|\right.\nonumber\\
&&\left.+\sqrt{k(N-k+1)}\left(|N,k\rangle|1\rangle\langle
N,k-1|\langle 0|\right.\right.\nonumber\\
&&\left.\left.+|N,k-1\rangle|0\rangle\langle N,k|\langle
1|\right)\right].
\end{eqnarray}
Further, substitution of the obtained expression into Eq.~(\ref{RG}) 
gives the operator $R^{(2)}_{G}$ in the form
\begin{equation}
R^{(2)}_G=\sum_{k=0}^{N+1}\frac{(N-k+1)}{(N+1)(N+2)}|N,k\rangle\langle
N,k|\otimes\openone_{\rm out}.
\end{equation}

\begin{figure}[!t!]
\centerline{\psfig{width=0.8\linewidth,angle=0,file=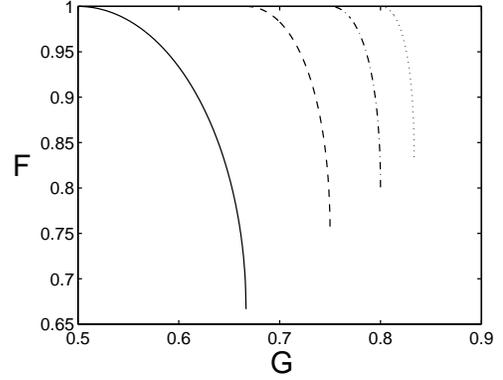}}
\caption{Optimal trade-off between the fidelities $F$ and $G$ for $N=1$ (solid curve), 
$N=2$ (dashed curve), $N=3$ (dotted-dashed curve) and $N=4$ (dotted curve) identical pure 
qubits.}
\label{fig1}
\end{figure}

In order to determine the optimal $\chi^{(2)}_{N}$ we have to find the maximum
eigenvalue and the corresponding eigenvector of the matrix $R^{(2)}_p$.
The matrix $R^{(2)}_p$ has a block diagonal structure with two one-dimensional blocks 
and $N$ two-dimensional blocks. The elements of the one-dimensional blocks are the 
eigenvalues $\lambda^{(0)}=\frac{(1-p)N+1}{(N+1)(N+2)}$ and $\lambda^{(N+1)}=1/(N+1)(N+2)$
with the characteristic subspaces spanned by the basis vectors $|N,0\rangle|1\rangle$ and 
$|N,N\rangle|0\rangle$, respectively. The two-dimensional blocks correspond to the invariant 
subspaces spanned by the basis vectors $\{|N,k-1\rangle|0\rangle, |N,k\rangle|1\rangle\}$, 
$k=1,2,\ldots,N$ and have the form ${\bf M}_{k}/(N+1)(N+2)$, where
\begin{eqnarray}
{\bf M}_{k}=
\left(\begin{array}{cc}
N-k+2 & p\sqrt{k(N-k+1)} \\
p\sqrt{k(N-k+1)} & N-k+1+p(2k-N)
\end{array}\right).\nonumber\\
\end{eqnarray}
The matrix ${\bf M}_{k}$ possesses two eigenvalues
\begin{equation}\label{lambda}
\mu^{(k)}_{1,2}=\frac{2N+3-pN}{2}-k(1-p)\pm\frac{\sqrt{(1+pN)^{2}-4pk(1-p)}}{2}
\end{equation}
from which one obtains the remaining eigenvalues of the matrix $R^{(2)}_{p}$ as 
$\lambda^{(k)}_{1,2}=\mu^{(k)}_{1,2}/(N+1)(N+2)$. The larger eigenvalue $\lambda^{(k)}_{1}$ 
is a decreasing function of $k$ attaining maximum for $k=1$. Obviously, 
$\lambda^{(1)}_{1}\geq\lambda^{(N+1)}$. Moreover, using in Eq.~(\ref{lambda}) the 
inequality $\sqrt{(1+Np)^{2}-4p(1-p)}\geq1-p$ following from the inequality $N\geq 1$ one can 
show that also $\lambda^{(1)}_{1}\geq\lambda^{(0)}$ holds and therefore $\lambda^{(1)}_{1}$ is 
the maximum eigenvalue of the matrix $R^{(2)}_{p}$. The eigenvalue is non-degenerate 
and its eigenvector determining the optimal map $\chi^{(2)}_{N}$ reads as
\begin{equation}\label{Nqubitmap1}
|\chi^{(2)}_{N}\rangle=\sqrt{N+1}\left(\alpha|N,0\rangle|0\rangle+
\beta|N,1\rangle|1\rangle\right),
\end{equation}
where $\alpha$ and $\beta$ are nonnegative real numbers satisfying the condition 
$\alpha^{2}+\beta^{2}=1$. On inserting $\chi^{(2)}_{N}=|\chi^{(2)}_{N}\rangle\langle\chi^{(2)}_{N}|$ 
into Eqs.~(\ref{FG}) one arrives after some algebra at the optimal fidelities
\begin{eqnarray}
F&=&\frac{1}{N+2}\left[\left(\sqrt{N}\alpha+\beta\right)^{2}+1\right],\label{qubitF}\\
G&=&\frac{1}{N+2}\left(N+\alpha^{2}\right)\label{qubitG}.
\end{eqnarray}
Expressing now the parameters $\alpha,\beta$ using Eq.~(\ref{qubitG}) and the 
normalization condition $\alpha^{2}+\beta^{2}=1$ and substituting the obtained 
formulas into Eq.~(\ref{qubitF}) we finally obtain the optimal fidelity trade-off 
for $N$ identical qubits (\ref{Nqubit}). The trade-off is depicted for several numbers 
of copies $N$  in Fig.~\ref{fig1}.  

The specific feature of the optimal map (\ref{Nqubitmap1}) is that it can be rewritten as the 
following coherent superposition of two maps:  
\begin{eqnarray}\label{Nqubitmap2}
|\chi^{(2)}_{N}\rangle&=&\sqrt{N+1}\left(\alpha'|N,0\rangle|0\rangle\right.\nonumber\\
&&\left.+\sqrt{\frac{2N}{N+1}}\beta'S_{N}|0\rangle^{\otimes N-1}|\Phi^{(2)}_{+}\rangle\right)
\end{eqnarray}
where $\alpha'=\alpha-\sqrt{N}\beta$, $\beta'=\sqrt{N+1}\beta$ and 
$|\Phi^{(2)}_{+}\rangle=(1/\sqrt{2})(|00\rangle+|11\rangle)$. The first 
map in the superposition is described by the vector $|N,0\rangle|0\rangle$ 
and corresponds to the choice $\alpha'=1$ ($\beta'=0$). Since in this case 
$F=G=\frac{N+1}{N+2}$ the map apparently realizes optimal estimation of a qubit from $N$ identical 
copies \cite{Massar_95}. The second map is obtained by choosing $\beta'=1$ ($\alpha'=0$) 
and it is represented by the second vector on the right hand side of Eq.~(\ref{Nqubitmap2}). 
It gives $F=1$ and $G=N/(N+1)$ which corresponds to optimal estimation of a qubit from 
$N-1$ copies while one copy is left intact by the map. 

\begin{figure}[!t!]
\centerline{\psfig{width=0.8\linewidth,angle=0,file=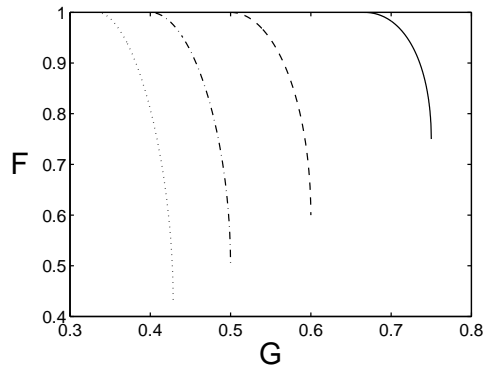}}
\caption{Optimal trade-off between the fidelities $F$ and $G$ for 
$N=2$ and $d=2$ (solid curve), $d=3$ (dashed curve), $d=4$ 
(dotted-dashed curve) and $d=5$ (dotted curve).}
\label{fig2}
\end{figure}
\section{Optimal partial measurement on $N$ qudits}\label{sec_2}

Interestingly, the fact that one can create a MDM as a coherent superposition of the 
two extreme maps is a general property of the MDMs that is valid not only for the present case of 
$N$ qubits, but it holds also for a single phase covariant qudit \cite{Mista_05}, 
two maximally entangled qudits \cite{Sacchi_06} or a single completely unknown qudit when 
\begin{equation}\label{quditmap}
|\chi^{(d)}_{1}\rangle=\sqrt{d}\left(\tilde{\alpha}|00\rangle+\tilde{\beta}|\Phi^{(d)}_{+}\rangle\right),
\end{equation}
where $|\Phi^{(d)}_{+}\rangle=(1/\sqrt{d})\sum_{j=0}^{d-1}|jj\rangle$ 
is the maximally entangled state of two qudits and $\tilde{\alpha},\tilde{\beta}\geq0$ satisfy the condition 
${\tilde\alpha}^{2}+{\tilde{\beta}}^{2}+2\tilde{\alpha}\tilde{\beta}/\sqrt{d}=1$.
Thus although we are not able to solve analytically the above optimization task 
of finding the MDM for $N$ qudits for a general $d$ we can use the superposition principle together 
with Eqs.~(\ref{Nqubitmap2}) and (\ref{quditmap}) to guess the desired optimal 
map to read 
\begin{equation}\label{Nquditmap1}
|\chi^{(d)}_{N}\rangle=\sqrt{D(N,d)}\left(\bar{\alpha}|0\rangle^{\otimes N+1}+
\bar{\beta}S_{N}|0\rangle^{\otimes N-1}|\Phi^{(d)}_{+}\rangle\right),
\end{equation}
where $\bar{\alpha},\bar{\beta}\geq0$ and 
\begin{equation}
{\bar{\alpha}}^{2}+\frac{2\bar{\alpha}\bar{\beta}}{\sqrt{d}}+\frac{N+d-1}{Nd}{\bar{\beta}}^{2}=1.
\end{equation}
In order to facilitate  the following calculations we rearrange 
the terms on the right hand side of the map (\ref{Nquditmap1}) and  rewrite it 
in the form:  
\begin{eqnarray}\label{Nquditmap2}
|\chi^{(d)}_{N}\rangle&=&\sqrt{D(N,d)}\left(\alpha|0\rangle^{\otimes N+1}+
\frac{\beta}{\sqrt{d-1}}\times\right.\nonumber\\
&\times&\left.\sum_{j=1}^{d-1}|N_{0}=N-1,N_{j}=1\rangle|j\rangle\right),
\nonumber\\
\end{eqnarray}
where $\alpha,\beta\geq0$ fulfill the condition $\alpha^{2}+\beta^{2}=1$ and where 
we have used the short hand notation $|N_{0}=N-1,N_{j}=1\rangle$ for a completely 
symmetric state of $N$ qudits containing $N-1$ qudits in the basis state $|0\rangle$ 
and a single qudit in the basis state $|j\rangle$. The fidelities $F$ and $G$ for this map 
can be again calculated with the help of Eq.~(\ref{FG}). Substituting Eq.~(\ref{Nquditmap2}) 
into Eq.~(\ref{FG}) and taking into account the symmetry of the projector $\Pi^{(d)}_{+,N+1}$ 
that implies $\left(\Pi^{(d)}_{+,N+1}\right)^{\rm T_N}=\left(\Pi^{(d)}_{+,N+1}\right)^{\rm T_{\rm out}}$ 
the problem of finding $F$ and $G$ reduces to the calculation of the following 
scalar products 
\begin{eqnarray}
&&A_{j}=\langle N_{0}=N-1,N_{j}=1|\langle 0|N_{0}=N,N_{j}=1\rangle,\nonumber\\
&&B_{j}=\langle N_{0}=N,N_{j}=1|0\rangle^{\otimes N}|j\rangle,\nonumber\\
&&C_{kj}=\nonumber\\
&&\langle N_{0}=N-1,N_{k}=1|\langle j|N_{0}=N-1,N_{k}=1,N_{j}=1\rangle,\nonumber\\
\end{eqnarray}
where we have used the short hand notation $|N_{0}=N-1,N_{k}=1,N_{j}=1\rangle$ for 
a completely symmetric state of $N+1$ qudits containing $N-1$ qudits in the basis 
state $|0\rangle$, a single qudit in the state $|k\rangle$ and a single qudit in 
the state $|j\rangle$. The scalar products can be easily evaluated 
using Eq.~(\ref{occupation}) as $A_{j}=\sqrt{N/(N+1)}$, $B_{j}=1/\sqrt{N+1}$ and 
$C_{jk}=\sqrt{(1+\delta_{jk})/(N+1)}$. Hence, one obtains 
\begin{eqnarray}\label{quditFG}
F&=&\frac{1}{N+d}\left[\left(\sqrt{N}\alpha+\sqrt{d-1}\beta\right)^{2}+1\right],\nonumber\\
G&=&\frac{1}{N+d}\left(N+\alpha^{2}\right).
\end{eqnarray}

Eliminating now the parameters $\alpha$ and $\beta$ from these equations using 
the same procedure as in the qubit case we arrive finally at the fidelity 
trade-off (\ref{Nqudit}). Although the found trade-off was not shown to 
be optimal here, there are several indications supporting our conjecture that 
it is really optimal. First, for $\beta=0$ we obtain $F=G=\frac{N+1}{N+d}$ 
using Eqs.~(\ref{quditFG}) and therefore these optimal fidelities satisfy 
our trade-off. Second, by putting $\alpha=\sqrt{N/(N+d-1)}$ and 
$\beta=\sqrt{(d-1)/(N+d-1)}$ one finds that $F=1$ and $G=N/(N+d-1)$ 
which means that also the second extreme case is fulfilled. 
Finally, for $d=2$ the trade-off reduces to the optimal trade-off for $N$ 
identical qubits (\ref{Nqubit}) while for $N=1$ it boils down to the optimal 
trade-off for a single completely unknown qudit \cite{Banaszek_01a}. The trade-off 
(\ref{Nqudit}) is depicted in Fig.~\ref{fig2} for $N=2$ and $d=2,3,4,5$.

\section{Conclusions}\label{sec_3}

In summary, in the present paper we have derived analytically 
the optimal trade-off between the mean estimation fidelity and the mean 
output fidelity for partial measurements on $N$ identical pure qubits. Furthermore, 
based on the structure of  the optimal map saturating the trade-off we have made 
a conjecture about the optimal fidelity trade-off for partial measurements on 
$N$ identical pure qudits. The obtained results provide an insight into the generic 
structure and properties of MDMs. The optimal partial measurements saturate 
the fundamental bound on conversion of quantum information onto classical information 
and may thus find applications in quantum communication and information processing.

\acknowledgments
The research has been supported by the research projects ``Measurement and 
Information in Optics,'' (MSM 6198959213) and Center of Modern Optics (LC06007) 
of the Czech Ministry of Education. Partial support by the SECOQC (IST-2002-506813) 
project of the sixth framework program of EU is also acknowledged.

\end{document}